\documentclass[conference]{IEEEtran}
\IEEEoverridecommandlockouts
\usepackage{cite}
\usepackage{amsmath,amssymb,amsfonts}
\usepackage{algorithmic}
\usepackage{graphicx}
\usepackage{textcomp}
\usepackage{xcolor}
\usepackage{url}
\def\BibTeX{{\rm B\kern-.05em{\sc i\kern-.025em b}\kern-.08em
    T\kern-.1667em\lower.7ex\hbox{E}\kern-.125emX}}
\begin{document}

\title{Trace Replay Simulation of MIT SuperCloud for Studying Optimal Sustainability Policies 
% Trace Replay Simulation of the MIT SuperCloud Dataset for Studying Optimal Policies for Sustainability
% Trace replay simulation of MIT SuperCloud dataset for studying optimal policies for sustainability
% \thanks{}
}

\author{\IEEEauthorblockN{Wesley Brewer}
\IEEEauthorblockA{\textit{Oak Ridge National Laboratory} \\
Oak Ridge, USA \\
brewerwh@ornl.gov}
\and
\IEEEauthorblockN{Matthias Maiterth}
\IEEEauthorblockA{\textit{Oak Ridge National Laboratory} \\
Oak Ridge, USA \\
maiterthm@ornl.gov}
\and
\IEEEauthorblockN{Damien Fay}
\IEEEauthorblockA{\textit{Hewlett Packard Enterprise} \\
Dublin, Ireland \\
damien.fay@hpe.com}
}

\maketitle

With the explosive growth of AI supercomputing, there is an increasing need for utility companies to accurately predict power demand of datacenters. For example, xAI recently built Colossus, a 200k GPU datacenter in Tennessee which will require 300MW at fully capacity~\cite{morales2025colossus}, Meta is building a 2.2 GW GPU datacenter in Louisiana, and OpenAI is building a 400k GPU datacenter ``Stargate'' in Texas powered by 1.2 GW~\cite{criddle2025openai}. 
Training LLMs on such machines typically induces large power swings, e.g., 20MW in milliseconds for the Frontier supercomputer. Accurate demand prediction is difficult for power companies facing such swings; efforts like DCFlex~\cite{dcflex} are seeking to help utilities better manage them.

To support efforts like DCFlex, we have developed a digital twin framework for modeling power and cooling of data centers called ExaDigiT\footnote{\url{https://exadigit.github.io}}~\cite{brewer2024digital}, which debuted at SC'24 and recently won a 2025 R\&D 100 award. It provides three components: (1) a resource allocator and power simulator (RAPS) to model the datacenter, (2) a cooling model to model the cooling infrastructure, and (3) a visual analytics module consisting of augmented/virtual reality, as well as a web-based dashboard. ExaDigiT has been used to develop digital twins of several supercomputers, including Frontier at Oak Ridge National Laboratory, Fugaku in Japan, LUMI in Finland, Adastra in France, Setonix in Australia, Marconi100 in Italy, and Lassen at Lawrence Livermore National Laboratory. ExaDigiT has evolved into a global open source community with over 150 members, ten supercomputing centers from around the world, and several industrial partners including HPE and NVIDIA. 

The RAPS module originally was designed to replay job traces for predicting total system power. It also estimates losses due to AC-DC rectification and voltage conversion by estimating power conversion efficiency, power usage effectiveness, as well as energy efficiency in terms of GFlops/watts. It has since been modified to support several scheduling algorithms that are able to match production Slurm scheduling capabilities~\cite{maiterth2025hpc}. Therefore, it can be used as as tool to study optimal scheduling policies, particularly in the context of energy efficiency.

Over the past year, we have been working to extend ExaDigiT to work with open telemetry datasets, including the PM100 dataset~\cite{antici2023pm100}, the Lassen dataset~\cite{patki2021monitoring}, and F-Data~\cite{antici2025f}, among others. Each dataset provides unique types of information for studying different problems. For example, the F-Data dataset contains labels about whether jobs are compute bound or memory bound, whereas the Lassen dataset contains total number of bytes sent into and out of each node, which is useful for studying network inter-job congestion and its relation to energy consumption. 

Recently, we are interested in exploring the modeling of cloud systems, specifically hyperscalers. Such systems introduce new challenges for modeling, especially in terms of heterogeneity and multi-tenancy. The MIT Supercloud dataset~\cite{samsi2021supercloud} provides rich information (e.g., includes power data at 100ms intervals, and includes CPU frequency, among others), such modeling could be used to develop a simulation, which can be used as a training environment for reinforcement learning. Such an environment may be used to train agent(s) to optimally schedule jobs for increasing throughput to reduce overall energy consumption. 

While this work is still in progress, its novelty lies in the fact that ExaDigiT can create a virtual cloud system, which can be used for testing optimal policies, scheduling strategies, incentive structures~\cite{solorzano2024toward}, virtual prototyping of hardware/software, and virtual benchmarking of speculative systems. 
In this work, we focus on extending ExaDigiT to support trace replay simulation and rescheduling of workloads on the MIT Supercloud TX-GAIA system~\cite{samsi2021supercloud} to serve as a training environment for reinforcement learning (RL), for which we have also developed an initial RL infrastructure.
% In this work, we focus on extending ExaDigiT to support trace replay simulation and rescheduling of workloads on the MIT Supercloud TX-GAIA system~\cite{samsi2021supercloud}, and we have developed an initial RL training environment.
% In this work, we focus on extending ExaDigiT to support trace replay simulation and rescheduling of workloads on the MIT Supercloud TX-GAIA system~\cite{samsi2021supercloud} to create a training environment for reinforcement learning optimization, and we have developed an initial RL training environment.

% \begin{figure}
%     \centering
%     \includegraphics[width=0.5\linewidth]{figs/exadigit-raps.png}
%     \caption{ExaDigiT RAPS architecture.}
%     \label{fig:exadigit}
% \end{figure}

\begin{figure}
    \centering
    \includegraphics[width=\linewidth]{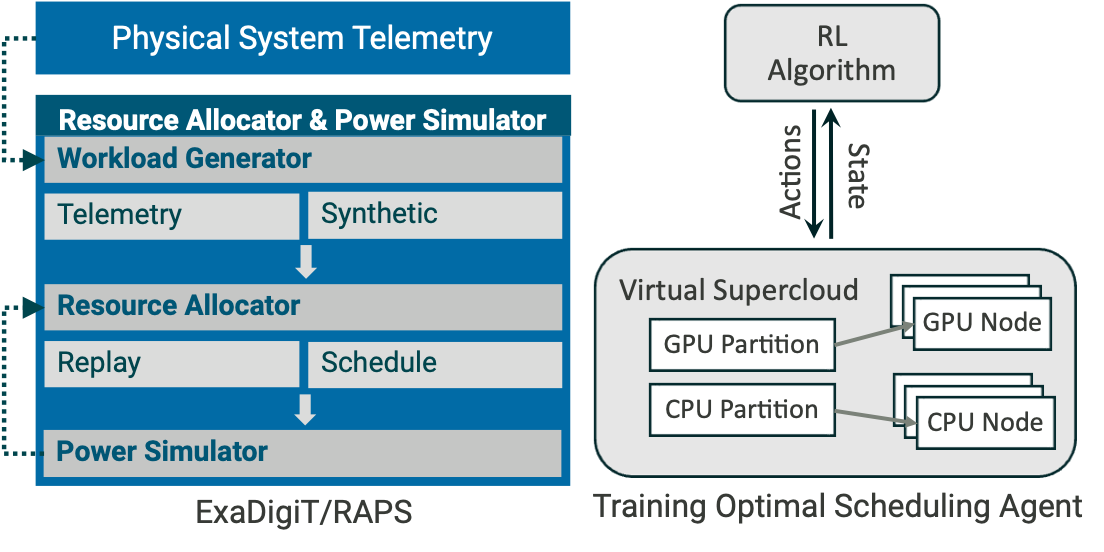}
    \caption{ExaDigiT/RAPS architecture (left). RL process (right).}
    \label{fig:exadigit}
\end{figure}

\begin{figure}
    \centering
    \includegraphics[width=\linewidth]{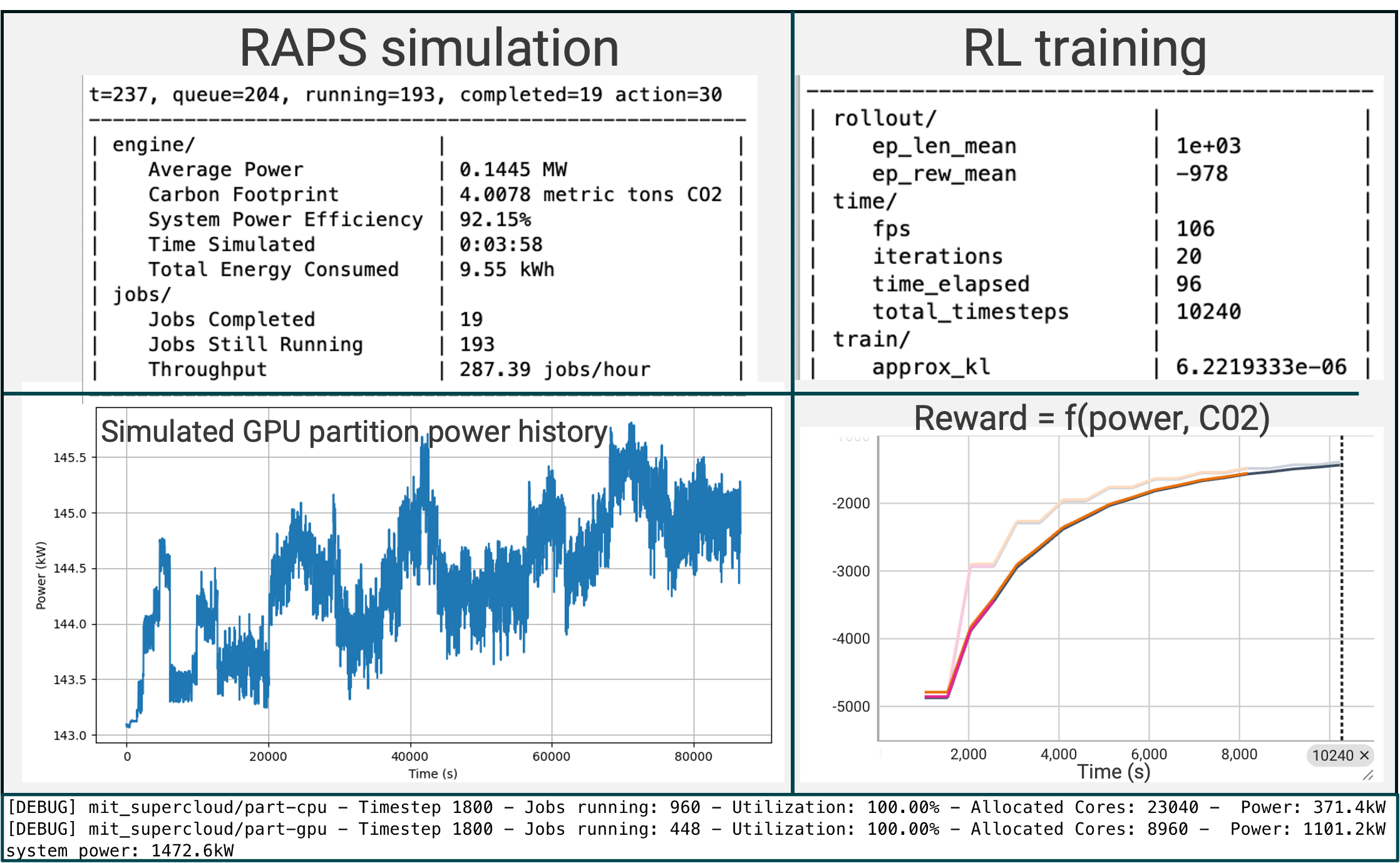}
    \caption{Preliminary integration of RL into RAPS. (Top-left) Simulation runtime stats showing job throughput and energy consumption. (Top-right) RL training metrics from PPO (average episodic reward improving over timesteps). (Bottom-right) System power history trace under RL-driven scheduling. (Bottom) Logs from peak utilization test with power prediction.}
    \label{fig:rl}
\end{figure}

Figure \ref{fig:exadigit} shows an architectural overview of ExaDigiT/RAPS, which either takes system telemetry traces, or can generate synthetic workloads using performance modeling tools, such as Calculon~\cite{isaev2023calculon} or PerfVec~\cite{li2024learning}. Traces are typically averaged over a specified trace quanta, which is usually on the order of seconds, e.g., 10s for the MIT Supercloud CPU telemetry, and 100ms for the GPU telemetry. RAPS steps forward in time, at a specified time delta, typically 1s, schedules jobs waiting in the queue, aggregates total power on all nodes, including rectification and voltage conversion losses, and tracks a large number of statistics, including total system power, energy conversion efficiency, throughput, system FLOPS, system utilization, etc. ExaDigiT has several built-in schedulers which can be invoked to either replay the system telemetry as scheduled, or can re-schedule the telemetry using different types of scheduling and backfill policies~\cite{maiterth2025hpc}. ExaDigiT/RAPS has also been used to virtually replace hardware, e.g., testing out ideas such as smart load-sharing rectifiers, or converting a system to run on medium voltage DC direct power~\cite{wojda2024dynamic}. Finally, RAPS can be used to model network congestion~\cite{holmen2024towards}. 
% , assuming network data is available for the system to be modeled. 

Reinforcement learning (RL) seeks to learn a \textit{policy} which maps \textit{states} to \textit{actions}, as shown in Fig.~\ref{fig:exadigit}. While traditional machine learning requires significant amounts of data to train the model, RL agents are trained by interacting with an environment, e.g. a simulation of the supercomputer. RL works by optimizing a \textit{reward} function by trial-and-error of actions to learn how the environment responds.
The \textit{action} space may represent scheduling decisions such as which job to dispatch, when to backfill, and where to place jobs on nodes or partitions, while the \textit{state} space captures the current queue, running jobs, available resources, and system-level metrics such as power consumption and throughput; the \textit{reward} function combines energy consumption, carbon footprint, and job throughput to guide energy-aware scheduling.
Fan et al.~\cite{fan2021deep} show a 45\% improvement in job slowdown, a measure of job response time (wait time + run time) divided by job run time, while our framework additionally predicts system power, energy, carbon footprint, and network congestion.
% Similar work in this area by Fan et al.~\cite{fan2021deep} show a 45\% improvement in job slowdown, which is a measure of job response time (wait time + run time) divided by job run time. 
% Fan et al.~\cite{fan2021deep} show a 45\% improvement in job slowdown, which is a measure of job response time (wait time + run time) divided by job run time, whereas our framework also models system power and carbon footprint.

The state of our project is: (1) developed an MIT Supercloud dataloader for RAPS, (2) extended RAPS to support heterogeneity and multi-tenancy, (3) enabled replay and rescheduling of jobs, and (4) integrated an OpenAI Gym-compatible interface for reinforcement learning experiments. Early Proximal Policy Optimization (PPO) training runs (Fig.~\ref{fig:rl}) based on Stable Baselines 3 show that RAPS can serve as an RL environment, with episodic rewards evolving over timesteps and metrics including throughput, carbon footprint, and energy consumption tracked. 
While validation of the RL agent and multi-tenant scheduler remains, these results demonstrate the framework’s potential for energy-aware scheduling.

\section*{Acknowledgment}
This research was sponsored by and used resources of the Oak Ridge Leadership Computing Facility (OLCF), which is a DOE Office of Science User Facility at the Oak Ridge National Laboratory (ORNL) supported by the U.S. Department of Energy under Contract No. DE-AC05-00OR22725. Also, we gratefully acknowledge the support of Albert Reuther at MIT Lincoln Laboratory for helping us to understand how to parse the MIT SuperCloud dataset.

\bibliographystyle{IEEEtran}
\bibliography{main}

\end{document}